\documentclass[runningheads]{llncs}
\usepackage[T1]{fontenc}
\usepackage{graphicx}
\usepackage{cite}
\usepackage{amsmath,amssymb,amsfonts}
\usepackage{algorithmic}
\usepackage{graphicx}
\usepackage{textcomp}
\def\BibTeX{{\rm B\kern-.05em{\sc i\kern-.025em b}\kern-.08em
    T\kern-.1667em\lower.7ex\hbox{E}\kern-.125emX}}

\usepackage{enumitem}
\usepackage{booktabs}
\usepackage[colorlinks=true,linkcolor=blue,anchorcolor=blue,citecolor=blue,urlcolor=blue]{hyperref}
\usepackage{bbding}
\usepackage{mwe}
\usepackage{hyperref}
\usepackage{float}
\usepackage{latexsym}
\usepackage{multirow}
\usepackage[table]{xcolor}
\usepackage{makecell}

\begin{document}
\title{Diff4MMLiTS: Advanced Multimodal Liver Tumor Segmentation via Diffusion-Based Image Synthesis and Alignment}
\titlerunning{Diff4MMLiTS for Multimodal Liver Tumor Segmentation}
\author{Shiyun Chen\inst{1\dagger} \and
Li Lin\inst{1,2\dagger} \and
Pujin Cheng\inst{1,2} \and
Zhicheng Jin\inst{3} \and
Jianjian Chen\inst{3} \and
Haidong Zhu\inst{3} \and
Kenneth K. Y. Wong\inst{2} \and
Xiaoying Tang\inst{1,4,*}}
\authorrunning{S. Chen et al.}

\renewcommand\thefootnote{}
\footnotetext{$^\dagger$ These authors contributed equally.}

\institute{Department of Electronic and Electrical Engineering, Southern University of Science and Technology, Shenzhen, China \\
\email{tangxy@sustech.edu.cn} \and
Department of Electrical and Electronic Engineering, The University of Hong Kong, Hong Kong SAR, China \and
Department of Radiology, Zhongda Hospital, Medical School, Southeast University, Nanjing, China \and
Jiaxing Research Institute, Southern University of Science and Technology, Jiaxing, China
}

\maketitle
\begin{abstract}
Multimodal learning has been demonstrated to enhance performance across various clinical tasks, owing to the diverse perspectives offered by different modalities of data. However, existing multimodal segmentation methods rely on well-registered multimodal data, which is unrealistic for real-world clinical images, particularly for indistinct and diffuse regions such as liver tumors. In this paper, we introduce Diff4MMLiTS, a four-stage multimodal liver tumor segmentation pipeline: pre-registration of the target organs in multimodal CTs; dilation of the annotated modality's mask and followed by its use in inpainting to obtain multimodal normal CTs without tumors; synthesis of strictly aligned multimodal CTs with tumors using the latent diffusion model based on multimodal CT features and randomly generated tumor masks; and finally, training the segmentation model, thus eliminating the need for strictly aligned multimodal data. Extensive experiments on public and internal datasets demonstrate the superiority of Diff4MMLiTS over other state-of-the-art multimodal segmentation methods.

\end{abstract}
\section{Introduction}
Liver tumor segmentation is crucial for clinical diagnosis and treatment. Previous studies have achieved notable results in unimodal liver tumor segmentation, confirming the effectiveness of deep learning techniques in this area \cite{3dunet, nnunet}. As different types of tumors exhibit distinct characteristics across various modalities, clinical radiologists often use multimodal CTs for more comprehensive diagnosis and more precise delineation of tumor boundaries \cite{zhang2023}. Integrating multimodal information significantly enhances segmentation performance \cite{maml, mmformer, panet}. However, compared to unimodal, the high annotation costs and data scarcity significantly constrain the exploration of multimodal segmentation methods.

Recent multimodal learning studies have underscored the importance of data alignment \cite{prior,llava}. Alignment is equally vital in multimodal medical image segmentation tasks. Public multimodal benchmarks like BraTS \cite{brats} and AutoPET \cite{autopet} either employ standardized anatomical templates for rigid alignment or are collected simultaneously, inherently ensuring rigid data alignment. Thus, most current multimodal tumor segmentation methods operate under the assumption of strict data alignment \cite{f2net, a2fseg}. In contrast, in real clinical settings, multimodal CTs often exhibit significant misalignment due to spatiotemporal differences in data acquisition. Although some registration methods based on traditional paradigms \cite{fsl} or deep learning \cite{cyclemorph, transmorph} have been proposed and achieved reasonable registration at the organ or tissue level, they still struggle to fully align the diffuse and indistinct tumor regions. This presents unavoidable noise for existing multimodal segmentation algorithms.

In recent years, due to the improvement of generative models, tumor segmentation based on synthesis has become a highly regarded research topic \cite{red,xpx}. Previous studies have demonstrated the effectiveness of using synthetic images to enhance liver tumor segmentation \cite{syntumor, difftumor,freetumor}. However, these methods are primarily limited to unimodal images and fail to achieve expected performance when directly applied to multimodal tasks. Nevertheless, these efforts provide two key insights: first, synthetic tumor samples can be used to alleviate the challenge of data scarcity and enhance model generalization; second, synthesizing rigorously aligned multimodal CTs can partially mitigate the current dependence of segmentation algorithms on well-registered multimodal data.

In this context, we present an innovative and effective multimodal segmentation framework named Diff4MMLiTS. Specifically, the process begins with preliminary registration and data preprocessing, during which the target organ is roughly aligned, but the tumor remains misaligned. Therefore, we employ the mask of the annotated modality, dilating it to ensure adequate coverage of the tumor across various modalities. Subsequently, this mask is utilized in the Normal CT Generator (NCG) module for inpainting, yielding the normal CT. Within the latent space, the Multimodal CT Synthesizer (MCS) module leverages the multimodal CT features and randomly generated tumor masks to synthesize strictly aligned multimodal CTs with tumors. The aforementioned synthesis process employs the common mask, yet each modality’s CT is synthesized according to its unique feature distribution mapping. Finally, the roughly aligned real data is combined with the fully aligned synthetic data to train the segmentation model. Our contributions are as follows:

\begin{figure*}[!t]
    \centering
    \includegraphics[width=0.95\textwidth]{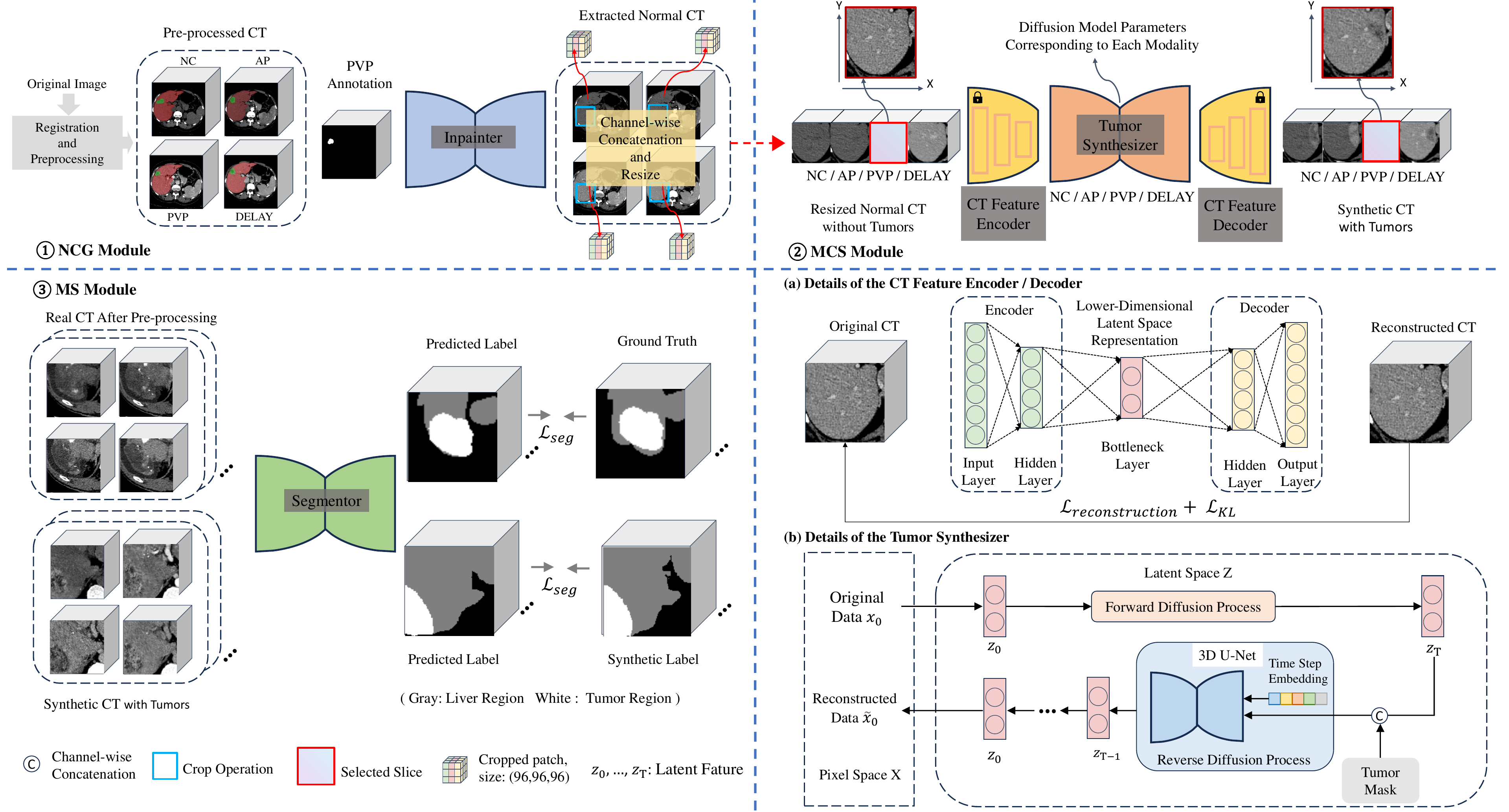}
    \vspace{-0.1cm}
    \caption{The architecture of Diff4MMLiTS. The NCG module uses the extended PVP mask to inpaint multimodal images to acquire normal CTs. The MCS module uses normal CTs to synthesize multimodal CTs. The MS module trains segmenter using real and synthetic data. MCS comprises two components: a 3D autoencoder consisting of a CT feature encoder and decoder, a tumor synthesizer based on a diffusion model.}
    \label{fig:fig1}
    \vspace{-5pt}
\end{figure*}

\begin{itemize}
\setlength{\itemsep}{0pt} 
\item An innovative multimodal liver tumor segmentation framework, Diff4MMLiTS, is proposed to effectively leverage unaligned multimodal liver tumor images.
\item The designed Inpainting-Synthesis-Segmentation pipeline leverages a diffusion model to synthesize a large number of perfectly aligned multimodal liver tumor images, thereby significantly improving segmentation performance.
\item Diff4MMLiTS is evaluated on both internal and external datasets, with experimental results demonstrating outstanding performance. 
\end{itemize}

\section{Method}
Diff4MMLiTS comprises three pivotal components: the NCG module, the MCS module, and the MS module. These components are trained sequentially, with each preceding stage’s output serving as input for the next. NCG provides the CT background for MCS to synthesize CTs with tumor. MCS then supplies MS with synthetic images to train the segmenter. The following subsections will provide a detailed elaboration. The overall framework is shown in Fig.~\ref{fig:fig1}. 

\subsection{Normal CT Generator (NCG) Module}
Since there is no publicly available paired multimodal normal CT dataset, NCG is used to generate multimodal normal CT as the basis for tumor synthesis. Inspired by Suvorov \cite{lama}, we employ an inpainting network based on Fast Fourier Convolution (FFC). To reduce computational load, we perform slice-by-slice inpainting of 2D slices within 3D CT scans. Given an image $x$ and a binary mask of the inpainting regions $m$, they are stacked as follows: 
\begin{equation}
x^{\prime}=\operatorname{stack}(x\odot (1-m),m),
\end{equation}
where $x\odot (1-m)$ denotes the masked image. The feed-forward inpainting network $f_\theta(\cdot)$ processes the input $x^{\prime}$ in a fully convolutional manner and generates the inpainted image $\hat{x}$ = $f_\theta(x^{\prime})$. Specifically, the image and mask pair serve as network inputs, which are downsampled into FFC residual blocks. During FFC processing, the input is divided into two parallel branches: a local branch, using traditional convolution, captures local information, while a global branch, based on real FFT, acquires global contextual information. These two outputs are then cross-fused and channel-wise concatenated to produce the final output.

We fine-tune the model using pre-trained parameters and followed the training strategy of Suvorov et al. \cite{lama}. For the fine-tuning of NCG, we employ original internal multimodal data to adapt it to the inpainting of multimodal CTs. After training, the inpainter is utilized to remove tumor foreground regions from multimodal CT with tumors. For each tumor region in the portal venous phase (PVP) mask, a 5×5 kernel is first employed for morphological closing, followed by a single iteration of dilation using a 3×3 kernel, to mitigate subjective noise and ensure maximal coverage of tumor regions across all modalities. NCG used the dilated PVP noise annotation as the mask to remove the foreground from labeled samples, ultimately producing a corresponding normal CT for each CT.

\subsection{Multimodal CT Synthesizer (MCS) Module}
MCS is divided into two stages: first, a autoencoder is trained to learn comprehensive latent features, including the CT feature encoder $\mathcal{E}$ and decoder $\mathcal{D}$; second, a tumor synthesizer based on a latent diffusion model (LDM) \cite{ldm} is trained to generate synthetic images. This latent feature-driven diffusion process enables efficient and reliable tumor generation within a compact latent space.

Specifically, the VQGAN-based autoencoder performs the encoding and reconstruction between the image $x_0$ and its latent representation $z_0$, giving $\tilde{x}_0=\mathcal{D}(z_0)=\mathcal{D}(\mathcal{E}(x_0))$. Since the universal autoencoding stage only requires training once and can be utilized repeatedly for multiple diffusion model trainings \cite{ldm}, we adapt the pre-trained autoencoder \cite{difftumor}. The diffusion process comprises a forward process and a reverse process. The forward diffusion process transforms the data distribution into a standard Gaussian distribution at each time step $t \in \begin{bmatrix} 1,\cdots,T \end{bmatrix}$  by the following process:
\begin{equation}
    q\left(\boldsymbol{z}_t \mid \boldsymbol{z}_0 \right)=\mathcal{N}\left(
    \boldsymbol{z}_t \mid \sqrt{\bar{\alpha}_{t}} \boldsymbol{z}_{0},\left(1-\bar{\alpha}_{t}\right) \boldsymbol{\epsilon}\right),
\end{equation}
where $\epsilon$ is the random noise, $\bar{\alpha}_{t}$ denotes the noise level.

The reverse generation process aims to gradually generate data samples from the noise by the following process:
\begin{equation}
    p\left(z_{0}\right)= p\left(z_{T}\right) \prod_{t=1}^{T} p\left(z_{t-1} \mid z_{t}, y\right),
\end{equation}
where $y$ is the tumor mask for conditioning information. During the inference stage, a coordinate within the liver region is randomly selected as the center point, and random values for the semi-axes $x$, $y$, and $z$ are assigned. An ellipsoidal mask approximating the tumor is then generated based on these parameters, followed by elastic deformation to produce the tumor mask $y$, indicating the location of the synthesized tumor. During the training stage, $\epsilon_{\theta}\left(\cdot \right)$ denotes the denoising encoder, the training objective of LDM can be written as:
\begin{equation}
    \mathcal{L}_{L D M}:=\mathbb{E}_{\mathcal{E}(x_0), \epsilon \sim \mathcal{N}(0,1), t}\left[\left\|\epsilon-\epsilon_{\theta}\left(z_{t}, y, t\right)\right\|_{2}^{2}\right].
\end{equation}

\subsection{Multimodal Segmentation (MS) Module}
\vspace{-0.1cm}
The segmenter is a U-shape network. During training, MS dynamically synthesizes CTs with tumors from inpainted CTs, thereby constructing a hybrid dataset that combines synthetic and real data. The training objective of the segmenter is to minimize the voxel classification error relative to the ground truth. The loss function $\mathcal{L}_{seg}$ is consist of the Dice loss $\mathcal{L}_{dice}$ and the cross-entropy loss $\mathcal{L}_{ce}$:
\begin{equation}
    \label{eq:segloss}
    \mathcal{L}_{seg} = \mathcal{L}_{dice} + \gamma \mathcal{L}_{ce},
\end{equation}
where $\gamma = 0.5$ is a factor for balancing the two loss contributions. During inference, if the input does not consist of all four modalities, certain modalities will be duplicated to match the input size required by the model. 

\section{Experiments and Results}
\subsection{Datasets} 
To the best of our knowledge, there is currently no publicly available well-aligned multimodal liver tumor CT dataset. Therefore, we construct an in-house dataset from the Second Affiliated Hospital of Southeast University, called mmLiTS. mmLiTS includes 45 sets of four-phase abdominal CTs with in-plane resolution of 512 × 512, spacing from 0.6445 mm to 0.9160 mm, slice counts ranging from 45 to 561, and thicknesses between 1.0 mm and 5.0 mm. Liver tumors are manually annotated by experienced radiologists, and the dataset is randomly split into training, validation, and test sets in a 3:1:1 ratio. Given that radiologists often rely on PVP images to identify tumor margins that are less discernible in other phases, annotations are provided for PVP. To further assess generalizability, we evaluate our method on the LiTS dataset \cite{lits}. As LiTS is a unimodal dataset without alignment issues, we replicate each image four times to match the input format required by our segmentation model.

\subsection{Implementation Details}
Diff4MMLiTS is implemented by PyTorch on a workstation equipped with NVI-DIA RTX 3090. We use nnUNet as the default backbone. To avoid over-fitting, online data augmentations, including random flipping, rotation, and color jittering, are applied. All methods are trained for 1000 epochs. The preprocessing involves clipping the images to a [-21, 189] window and normalizing them to zero mean and unit standard deviation, followed by random cropping of 96×96×96 patches as input. The postprocessing filters the detection results by retaining only tumors located within the liver region. 

\begin{table}[!t]
    \centering
    \caption{Quantitative comparisons of Diff4MMLiTS with state-of-the-art methods on mmLiTS. The best results are highlighted in bold.}
    \renewcommand\arraystretch{1.2}
  \setlength{\tabcolsep}{1.8mm}
  \renewcommand\theadgape{\Gape[1.8mm][0mm]}
  \resizebox{0.75\linewidth}{!}{
        \begin{tabular}{l|cccc}
            \Xhline{0.6pt}
            Methods & DSC(\%) & JAC(\%) & SE(\%) & PRE(\%)  \\
            \hline
            16'U-Net \cite{3dunet}  & 75.46 & 61.29 & 68.12 & 87.69  \\ 
            16'V-Net \cite{vnet} & 75.97 & 62.70 & \textbf{81.78} & 74.40  \\
            18'AttentionUNet \cite{attentionnet} & 67.80 & 53.75 & 65.25 & 77.72  \\
            18'nnUNet \cite{nnunet} &76.34 &62.77 &69.84 &86.47  \\
            22'SwinUNETR \cite{swinunetr} &77.28 &63.41 &71.72 &86.02\\
            \hline
            18'SegResNet \cite{segresnet} & 77.73 & 64.05 & 75.29 & 81.87  \\
            21'MAML \cite{maml} & 76.06 & 63.08 & 67.92 & \textbf{91.27}  \\
            22'mmformer \cite{mmformer} & 77.32 & 64.32 & 78.76 & 77.94 \\
            23'A2FSeg \cite{a2fseg} &69.74 &55.90 &61.93 &88.66 \\
            24'PA-Net \cite{panet} &76.56 &63.09 &74.38 &82.23\\
            \hline
            Diff4MMLiTS & \textbf{79.02} & \textbf{66.33} & 74.47 & 85.24 \\
            \Xhline{0.6pt}
        \end{tabular}
    }
    \label{tab:table1}  
\end{table}

\begin{figure*}[!t]
    \centering
    \includegraphics[width=1\textwidth]{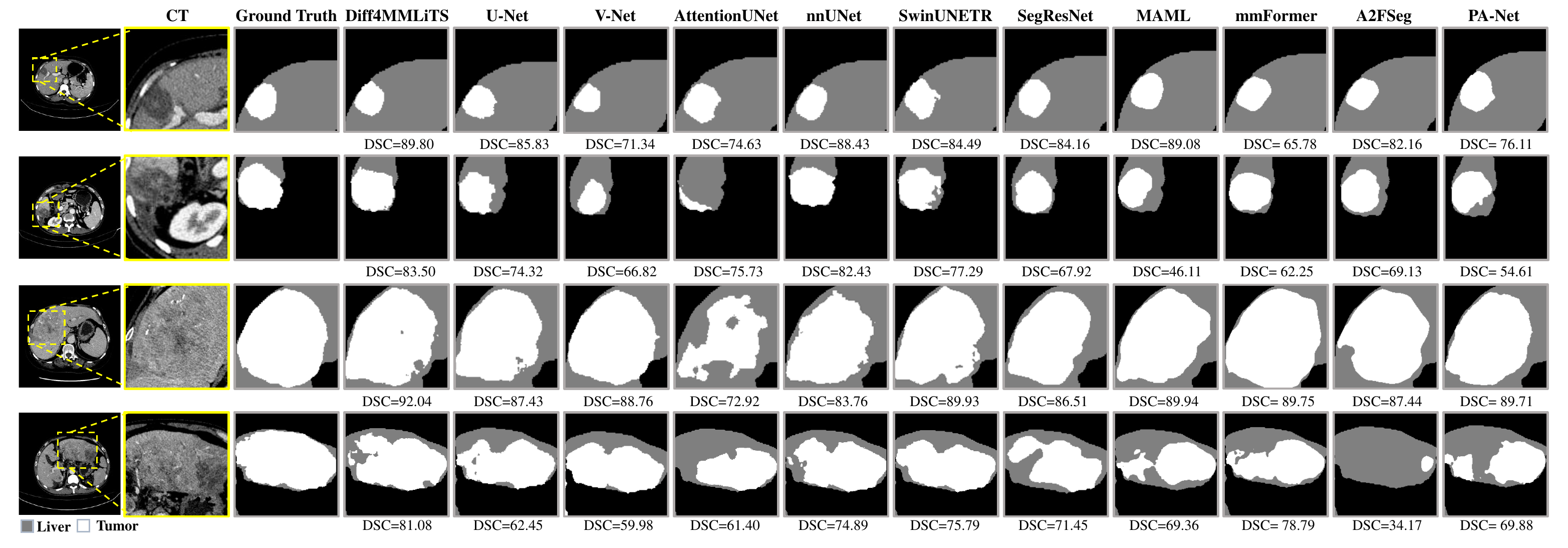}
    \vspace{-0.6cm}
    \caption{Qualitative visualization results.}
    \label{fig:fig2}
    \vspace{-5pt}
\end{figure*}

\subsection{Overall Performance}
To evaluate our method, we compare Diff4MMLiTS with various general and multimodal segmentation methods. Given that achieving precise liver segmentation is relatively straightforward, we focus specifically on the segmentation of intrahepatic tumors. The comparison results are presented in Table \ref{tab:table1}, with visualizations shown in Fig. \ref{fig:fig2}.

Compared to the U-Net and its variants, our method enhances the DSC by 1.74\% over the second-best SwinUNETR. When compared with other multimodal medical image segmentation models, our method enhances DSC by 1.29\% over the second best SegResNet. Moreover, Diff4MMLiTS significantly reduces false positives and false negatives, as reflected in the improved SE and PRE. This indicates enhanced tumor detection capability and a diminished likelihood of missed or incorrect detections. Overall, our proposed method has excellent superiority and achieves the best performance in the multimodal liver tumor segmentation task.

\subsection{Generalizability and Adaptability of Diff4MMLiTS}
We evaluate the proposed method on a publicly available external dataset to assess its out-of-distribution generalization capability, demonstrating that the model trained with Diff4MMLiTS performs effectively without retraining. All methods are trained on mmLiTS and tested on lesion samples from LiTS. The comparison results are shown in Table~\ref{tab:table2}. Compared to nnUNet fully trained on real data, Diff4MMLiTS with the synthesis strategy achieves a 16.12\% improvement in DSC. This implies that training models with such reliable synthetic images can effectively mitigate the risk of overfitting to in-distribution samples, thereby enhancing their ability to generalize to out-of-distribution samples.

To further assess the adaptability of Diff4MMLiTS, we integrate three different backbone models into the MS module, with the results summarized in Table~\ref{tab:table3}. The results demonstrate that our framework adapts well to these backbones, yielding notable performance gains. Compared to models trained solely on real data, those using the hybrid training strategy achieve DSC improvements of 1.95\%, 6.58\%, and 2.68\%, respectively. These findings highlight both the flexibility of our framework across different backbones and the effectiveness of the hybrid training strategy in enhancing segmentation performance.

\begin{table}[!t]
    \centering
    \caption{Comparison of experimental results on LiTS. The best results are highlighted in bold.}
    \renewcommand\arraystretch{1.2}
  \setlength{\tabcolsep}{2mm}
  \renewcommand\theadgape{\Gape[1.8mm][0mm]}
  \resizebox{0.75\linewidth}{!}{
        \begin{tabular}{l|cccc}
            \Xhline{0.6pt}
            Methods & DSC(\%) & JAC(\%) & SE(\%) & PRE(\%)  \\
            \hline
            nnUNet (real tumors) &41.63 &27.37 &55.51 &35.56  \\
        Diff4MMLiTS & \textbf{57.75} & \textbf{42.99} & \textbf{59.41} & \textbf{64.59}  \\
            \Xhline{0.6pt}
        \end{tabular}
    }
    \label{tab:table2}  
\end{table}

\begin{table}[!t]
  \centering
  \caption{Results of Diff4MMLiTS equipped with different backbone models. The best results are highlighted in bold.}
  \renewcommand\arraystretch{1.2}
  \setlength{\tabcolsep}{0.6mm}
  \renewcommand\theadgape{\Gape[1.8mm][0mm]}
  \resizebox{0.75\linewidth}{!}{
      \begin{tabular}{l|cccc}
            \Xhline{0.6pt}
            \multirow{1}{*}{\makecell[l]{Methods}}             
            & DSC(\%) & JAC(\%) & SE(\%) & PRE(\%)\\
            \hline
            U-Net \cite{3dunet}  & 75.46 & 61.29 & 68.12 & 87.69  \\ 
            U-Net w/ Diff4MMLiTS &77.41 &64.25 &72.22 &85.17 \\
            \hline
            AttentionUNet \cite{attentionnet} & 67.80 & 53.75 & 65.25 & 77.72 \\
            AttentionUNet w/ Diff4MMLiTS & 74.38 & 60.40 &67.37 &84.58 \\
            \hline
            nnUNet \cite{nnunet} &76.34 &62.77 &69.84 &86.47 \\
 nnUNet w/ Diff4MMLiTS & \textbf{79.02} & \textbf{66.33} & \textbf{74.47} & \textbf{85.24} \\
            \Xhline{0.6pt}
        \end{tabular}
    }
    \label{tab:table3}  
\end{table}

\subsection{Effectiveness of Multimodal Synthesis Strategy}
\vspace{-0.1cm}
As shown in Table~\ref{tab:table4}, we evaluate the effectiveness of our synthesis strategy on both multimodal and unimodal segmentation methods, using nnUNet as the segmentation backbone in all settings. Our proposed method achieves superior performance using only a small subset of multimodal data. Notably, with just 10\% of the data (real samples and inpainted multimodal CTs from three patients), our method outperforms the unimodal model trained on fully annotated data by 2.63\% in DSC. Furthermore, compared to fully supervised multimodal models summarized in Table~\ref{tab:table1}, our method achieves comparable performance using only 70\% of the training data. These results demonstrate the efficiency and effectiveness of Diff4MMLiTS in enhancing liver tumor segmentation.

\begin{table}[!t]
\caption{Results using only a portion of real diseased CT and corresponding normal CT training. Note that uni is short for unimodal and multi is short for multimodal.}
\label{tab:table4}  
\centering
\renewcommand\arraystretch{1.2}
  \setlength{\tabcolsep}{1.3mm}
  \renewcommand\theadgape{\Gape[1.8mm][0mm]}
  \resizebox{0.7\linewidth}{!}{
\begin{tabular}{cc|cccc}
\Xhline{0.6pt}
\multicolumn{2}{c|}{Ratio(diseased/normal)} & DSC(\%) & JAC(\%) & SE(\%) & PRE(\%)  \\ 
\hline
uni &100\% (29/29)  &71.24 &56.08 &64.31 &85.11\\
\hline
\multirow{5}{*}{multi} & 10\% (3/3) &73.87 &59.58 &76.40 &73.13  \\
&30\% (9/9) &75.26 &60.53 &70.24 &84.14  \\
&50\% (15/15) &76.97 &63.45 &70.25 &87.43  \\
&70\% (20/20) &78.49 &65.47 &75.66 &82.47 \\
&100\% (29/29) &79.02 &66.33 &74.47 &85.24 \\
\Xhline{0.6pt}
\end{tabular}
}
\end{table}

\begin{table}[!t]
  \centering
  \caption{Results of Diff4MMLiTS equipped with different inpainting or synthesis methods. \$ and \# denotes the replacement of the inpainter and synthesizer, respectively. The best results are highlighted in bold.}
  \renewcommand\arraystretch{1.2}
  \setlength{\tabcolsep}{2.7mm}
  \renewcommand\theadgape{\Gape[1.8mm][0mm]}
  \resizebox{0.7\linewidth}{!}{
      \begin{tabular}{l|cccc}
            \Xhline{0.6pt}
            Methods & DSC(\%) & JAC(\%) & SE(\%) & PRE(\%)  \\ 
            \hline
            Median (\$) & 74.79 & 60.75 &\textbf{79.43} &73.37 \\
            CUT (\#) &77.64  &63.92  &74.78  & 84.62  \\
            Diff4MMLiTS & \textbf{79.02} & \textbf{66.33} & 74.47 & \textbf{85.24}\\
            \Xhline{0.6pt}
        \end{tabular}
    }
    \label{tab:table5}  
\end{table}

\subsection{Contribution of Different Modules to the Framework}
\vspace{-0.1cm}
To comprehensively validate the importance of each module, we explore alternative designs. Specifically, we replace the inpainter in NCG with a median filter. For MCS, we replace the LDM in MCS with CUT \cite{cutgan}, which constructs the source domain by concatenating normal CTs with randomly generated tumor masks and learns a mapping to the target domain of real tumor CTs. The impact of these alternatives on overall performance is shown in Table~\ref{tab:table5}. The results indicate that the framework with the original inpainter and synthesizer achieves the best performance. In comparison, CTs generated by the median filter introduce noise and disrupt the original details, thereby affecting the quality of subsequent tumor synthesis. Additionally, the tumor images synthesized by CUT are overly smooth, making the complementary features between multimodal tumors less pronounced, which, in turn, affects the performance of the segmentation model. In contrast, the inpainter and synthesizer proposed in our original framework not only preserve background details, but also generate reliable multimodal liver tumors, playing a crucial role in the liver tumor segmentation task.

\section{Conclusion}
\vspace{-0.2cm}
This paper presents a novel multimodal liver tumor segmentation framework, named Diff4MMLiTS. Diff4MMLiTS is a diffusion-based pipline comprising of a Normal CT Generator, a Multimodal CT Synthesizer and a Multimodal Segmenter. Extensive experiments demonstrate the superiority of our framework. A potential future direction is to employ an all-in-one diffusion model paradigm for multimodal tumor synthesis while exploring advanced network designs and multimodal fusion strategies to further enhance performance.

\section{Acknowledgments}
\vspace{-0.2cm}
This study was supported by the National Key Research and Development Program of China (2023YFC2415400); the National Natural Science Foundation of China (T2422012, 62071210); the Guangdong Basic and Applied Basic Research (2024B1515020088); the Shenzhen Science and Technology Program (RCYX2021\\
0609103056042); the High Level of Special Funds (G030230001, G03034K003).

\bibliographystyle{splncs04}
\bibliography{refs}

@article{lama,
  title={Resolution-robust Large Mask Inpainting with Fourier Convolutions},
  author={Suvorov, Roman and Logacheva, Elizaveta and Mashikhin, Anton and Remizova, Anastasia and Ashukha, Arsenii and Silvestrov, Aleksei and others},
  journal={Winter Conference on Applications of Computer Vision (WACV)},
  year={2021},
  pages={3172-3182},
  url={https://api.semanticscholar.org/CorpusID:237513361}
}

@inproceedings{ldm,
  title={High-resolution image synthesis with latent diffusion models},
  author={Rombach, Robin and Blattmann, Andreas and Lorenz, Dominik and Esser, Patrick and Ommer, Bj{\"o}rn},
  booktitle={Proceedings of the IEEE/CVF conference on computer vision and pattern recognition},
  pages={10684--10695},
  year={2022}
}

@article{freetumor,
  title={FreeTumor: Advance Tumor Segmentation via Large-Scale Tumor Synthesis},
  author={Wu, Linshan and Zhuang, Jiaxin and Ni, Xuefeng and Chen, Hao},
  journal={arXiv preprint arXiv:2406.01264},
  year={2024}
}

@inproceedings{difftumor,
  title={Towards generalizable tumor synthesis},
  author={Chen, Qi and Chen, Xiaoxi and Song, Haorui and Xiong, Zhiwei and Yuille, Alan and Wei, Chen and Zhou, Zongwei},
  booktitle={Proceedings of the IEEE/CVF Conference on Computer Vision and Pattern Recognition},
  pages={11147--11158},
  year={2024}
}

@inproceedings{syntumor,
  title={Label-free liver tumor segmentation},
  author={Hu, Qixin and Chen, Yixiong and Xiao, Junfei and Sun, Shuwen and Chen, Jieneng and Yuille, Alan L and Zhou, Zongwei},
  booktitle={Proceedings of the IEEE/CVF Conference on Computer Vision and Pattern Recognition},
  pages={7422--7432},
  year={2023}
}

@article{vnet,
  title={V-Net: Fully Convolutional Neural Networks for Volumetric Medical Image Segmentation},
  author={Milletari, Fausto and Navab, Nassir and Ahmadi, Seyed-Ahmad},
  journal={Fourth International Conference on 3D Vision},
  year={2016},
  pages={565-571},
  url={https://api.semanticscholar.org/CorpusID:206429151}
}

@article{attentionnet,
  title={Attention U-Net: Learning Where to Look for the Pancreas},
  author={Oktay, Ozan and Schlemper, Jo and Folgoc, Loic Le and Lee, Matthew and Heinrich, Mattias and Misawa, Kazunari and others},
  journal={ArXiv},
  year={2018},
  volume={abs/1804.03999},
  url={https://api.semanticscholar.org/CorpusID:4861068}
}

@article{nnunet,
  title={nnU-Net: a self-configuring method for deep learning-based biomedical image segmentation},
  author={Isensee, Fabian and Jaeger, Paul F and Kohl, Simon AA and Petersen, Jens and Maier-Hein, Klaus H},
  journal={Nature methods},
  volume={18},
  number={2},
  pages={203--211},
  year={2021},
  publisher={Nature Publishing Group}
}

@inproceedings{swinunetr,
  title={Swin unetr: Swin transformers for semantic segmentation of brain tumors in mri images},
  author={Hatamizadeh, Ali and Nath, Vishwesh and Tang, Yucheng and Yang, Dong and Roth, Holger R and Xu, Daguang},
  booktitle={International MICCAI brainlesion workshop},
  pages={272--284},
  year={2021},
  organization={Springer}
}

@inproceedings{segresnet,
  title={3D MRI brain tumor segmentation using autoencoder regularization},
  author={Andriy Myronenko},
  booktitle={BrainLes@MICCAI},
  year={2018},
  url={https://api.semanticscholar.org/CorpusID:53104235}
}

@inproceedings{maml,
  title={Modality-aware mutual learning for multi-modal medical image segmentation},
  author={Zhang, Yao and Yang, Jiawei and Tian, Jiang and Shi, Zhongchao and Zhong, Cheng and Zhang, Yang and He, Zhiqiang},
  booktitle={International Conference on Medical Image Computing and Computer-Assisted Intervention},
  pages={589--599},
  year={2021},
  organization={Springer}
}

@article{panet,
  title={PA-Net: A phase attention network fusing venous and arterial phase features of CT images for liver tumor segmentation},
  author={Liu, Zhenbing and Hou, Junfeng and Pan, Xipeng and Zhang, Ruojie and Shi, Zhenwei},
  journal={Computer Methods and Programs in Biomedicine},
  volume={244},
  pages={107997},
  year={2024},
  publisher={Elsevier}
}

@inproceedings{a2fseg,
  title={A2FSeg: Adaptive Multi-modal Fusion Network for Medical Image Segmentation},
  author={Wang, Zirui and Hong, Yi},
  booktitle={International Conference on Medical Image Computing and Computer-Assisted Intervention},
  pages={673--681},
  year={2023},
  organization={Springer}
}

@inproceedings{mmformer,
  title={mmformer: Multimodal medical transformer for incomplete multimodal learning of brain tumor segmentation},
  author={Zhang, Yao and He, Nanjun and Yang, Jiawei and Li, Yuexiang and Wei, Dong and Huang, Yawen and others},
  booktitle={International Conference on Medical Image Computing and Computer-Assisted Intervention},
  pages={107--117},
  year={2022},
  organization={Springer}
}

@article{lits,
  title={The liver tumor segmentation benchmark (lits)},
  author={Bilic, Patrick and Christ, Patrick and Li, Hongwei Bran and Vorontsov, Eugene and Ben-Cohen, Avi and Kaissis, Georgios and others},
  journal={Medical Image Analysis},
  volume={84},
  pages={102680},
  year={2023},
  publisher={Elsevier}
}

@article{prior,
  title={PRIOR: Prototype Representation Joint Learning from Medical Images and Reports},
  author={Cheng, Pujin and Lin, Li and Lyu, Junyan and Huang, Yijin and Luo, Wenhan and Tang, Xiaoying},
  journal={International Conference on Computer Vision (ICCV)},
  year={2023},
  pages={21304-21314},
  url={https://api.semanticscholar.org/CorpusID:260125025}
}

@article{cyclemorph,
  title={CycleMorph: Cycle Consistent Unsupervised Deformable Image Registration},
  author={Kim, Boah and Kim, Dong Hwan and Park, Seong Ho and Kim, Jieun and Lee, June-Goo and Ye, Jong Chul},
  journal={Medical Image Analysis},
  year={2020},
  volume={71},
  pages={
          102036
        },
  url={https://api.semanticscholar.org/CorpusID:221112297}
}

@article{transmorph,
  title={TransMorph: Transformer for unsupervised medical image registration},
  author={Chen, Junyu and Frey, Eric C and He, Yufan and Segars, William P and Li, Ye and Du, Yong},
  journal={Medical Image Analysis},
  year={2021},
  volume={82},
  pages={
          102615
        },
  url={https://api.semanticscholar.org/CorpusID:263478016}
}

@article{llava,
  title={Llava-med: Training a large language-and-vision assistant for biomedicine in one day},
  author={Li, Chunyuan and Wong, Cliff and Zhang, Sheng and Usuyama, Naoto and Liu, Haotian and Yang, Jianwei and others},
  journal={Advances in Neural Information Processing Systems},
  volume={36},
  year={2024}
}

@article{f2net,
  title={Flexible Fusion Network for Multi-Modal Brain Tumor Segmentation},
  author={Yang, Hengyi and Zhou, Tao and Zhou, Yi and Zhang, Yizhe and Fu, Huazhu},
  journal={IEEE Journal of Biomedical and Health Informatics},
  year={2023},
  volume={27},
  pages={3349-3359},
  url={https://api.semanticscholar.org/CorpusID:258438360}
}

@article{brats,
  title={The Multimodal Brain Tumor Image Segmentation Benchmark (BRATS)},
  author={Menze, Bjoern H and Jakab, Andras and Bauer, Stefan and Kalpathy-Cramer, Jayashree and Farahani, Keyvan and Kirby, Justin and others},
  journal={IEEE Transactions on Medical Imaging},
  year={2015},
  volume={34},
  pages={1993-2024},
  url={https://api.semanticscholar.org/CorpusID:1739295}
}

@article{autopet,
  title={A whole-body FDG-PET/CT Dataset with manually annotated Tumor Lesions},
  author={Gatidis, Sergios and Hepp, Tobias and Fr{\"u}h, Marcel and La Foug{\`e}re, Christian and Nikolaou, Konstantin and Pfannenberg, Christina and others},
  journal={Scientific Data},
  year={2022},
  volume={9},
  url={https://api.semanticscholar.org/CorpusID:252696059}
}

@article{zhang2023,
  title={Multi-modal tumor segmentation with deformable aggregation and uncertain region inpainting},
  author={Zhang, Yue and Peng, Chengtao and Tong, Ruofeng and Lin, Lanfen and Chen, Yen-Wei and Chen, Qingqing and Hu, Hongjie and Zhou, S Kevin},
  journal={IEEE Transactions on Medical Imaging},
  volume={42},
  number={10},
  pages={3091--3103},
  year={2023},
  publisher={IEEE}
}

@inproceedings{3dunet,
  title={3D U-Net: learning dense volumetric segmentation from sparse annotation},
  author={{\c{C}}i{\c{c}}ek, {\"O}zg{\"u}n and Abdulkadir, Ahmed and Lienkamp, Soeren S and Brox, Thomas and Ronneberger, Olaf},
  pages={424--432},
  year={2016},
  organization={Springer}
}

@inproceedings{red,
  title={Red-GAN: Attacking class imbalance via conditioned generation. Yet another medical imaging perspective.},
  author={Qasim, Ahmad B and Ezhov, Ivan and Shit, Suprosanna and Schoppe, Oliver and Paetzold, Johannes C and Sekuboyina, Anjany and others},
  booktitle={Medical imaging with deep learning},
  pages={655--668},
  year={2020},
  organization={PMLR}
}

@inproceedings{cutgan,
  title={Contrastive learning for unpaired image-to-image translation},
  author={Park, Taesung and Efros, Alexei A and Zhang, Richard and Zhu, Jun-Yan},
  booktitle={Computer Vision--ECCV 2020},
  pages={319--345},
  year={2020},
  organization={Springer}
}

@article{xpx,
  title={LF-SynthSeg: Label-Free Brain Tissue-Assisted Tumor Synthesis and Segmentation},
  author={Xu, Pengxiao and Lyu, Junyan and Lin,  Li and Cheng, Pujin and Tang, Xiaoying},
  journal={IEEE Journal of Biomedical and Health Informatics},
  year={2024},
  publisher={IEEE}
}

@article{fsl,
  title={Fsl},
  author={Jenkinson, Mark and Beckmann, Christian F and Behrens, Timothy EJ and Woolrich, Mark W and Smith, Stephen M},
  journal={Neuroimage},
  volume={62},
  number={2},
  pages={782--790},
  year={2012},
  publisher={Elsevier}
}

\end{document}